\documentclass[journal=jpclcd,manuscript=letter]{achemso}
\pdfoutput=1
\usepackage{graphics}
\usepackage{bm}
\usepackage{psfrag}
\usepackage{amssymb}
\usepackage{subcaption}

\title{Disentangling electronic and vibrational coherence in the Phycocyanin-645 light-harvesting complex}
\author{G.H. Richards}
\affiliation{Centre for Atom Optics and Ultrafast Spectroscopy,Swinburne University of Technology, Victoria 3122, Australia}

\author{K.E. Wilk}
\author{P.M.G. Curmi}
\affiliation{School of Physics and Centre for Applied Medical Research, St Vincents Hospital, The University of New South Wales, Sydney, New South Wales 2052, Australia}

\author{J.A. Davis}
\email{JDavis@swin.edu.au}
\affiliation{Centre for Atom Optics and Ultrafast Spectroscopy, and ARC Centre of Excellence for Coherent X-Ray Science, Swinburne University of Technology, Victoria 3122, Australia}

\begin{document}

\footnote{Corresponding author: jdavis@swin.edu.au}
\begin{abstract}

Energy transfer between chromophores in photosynthesis proceeds with near unity quantum efficiency. Understanding the precise mechanisms of these processes is made difficult by the complexity of the electronic structure and interactions with different vibrational modes. Two-dimensional spectroscopy has helped resolve some of the ambiguities and identified quantum effects that may be important for highly efficient energy transfer. Many questions remain, however, including whether the coherences observed are electronic and/or vibrational in nature and what role they play.  We utilise a two-colour four-wave mixing experiment with control of the wavelength and polarization to selectively excite specific coherence pathways.  For the light-harvesting complex PC645, from cryptophyte algae, we reveal and identify specific contributions from both electronic and vibrational coherences and determine an excited state structure based on two strongly-coupled electronic states and two vibrational modes. Separation of the coherence pathways also uncovers the complex evolution of these coherences and the states involved.

\end{abstract}


keywords: Ultrafast spectroscopy, cryptophyte, coherence, quantum coupling.
\newpage

Coherent multi-dimensional electronic spectroscopy typically utilises a series of femtosecond laser pulses to understand dynamics and relaxation pathways in complex systems. The broad spectral bandwidth of femtosecond laser pulses allows many different transitions and pathways to be excited and probed simultaneously.  The relaxation pathways and dynamics can then be identified in a 2D or 3D spectrum that correlates the state of the system in the different time periods between pulses. This type of technique has been immensely successful in NMR and IR spectroscopy for decades\cite{Tokmakoff2003,Noda1993} and more recently in visible/electronic spectroscopy\cite{Brixner2005,Cho2008}, where it has been used to identify coherent superpositions of excited states in photosynthetic light harvesting complexes that remain coherent for hundreds of femtoseconds\cite{Collini2010,Engel2007,Panitchayangkoon2010}. The original interpretation of these results attributed the coherences to superpositions of electronic eigenstates of the system, which triggered a large amount of research towards redeveloping theories of excitation energy transfer\cite{Ishizaki2009JCP,Ishizaki2009PNAS,AspuruGuzik2009,Plenio2008,Nazir2009,Castro2011,Jang2008}.  
 
There remain, however, several questions that need to be answered experimentally.  Among the most debated is whether the coherent superpositions observed in the multidimensional spectroscopy experiments are due to superpositions of electronic, vibrational or vibronic states, and if this has any bearing on the energy transfer mechanisms and efficiency. Each type of coherence is possible but separating the different contributions in 2D spectra can be difficult, particularly where there are many broad and overlapping transitions, as is the case in light harvesting complexes. Several means of identifying the nature of such coherences have been reported, including comparing data obtained from rephasing and non-rephasing pulse orderings\cite{Turner2011} and using polarization control of the individual pulses \cite{Westenhoff2012,SchlauCohen2012,Read2009} to selectively excite different signal pathways. There are, however,  limitations and challenges to each of these solutions which are exacerbated in systems such as light harvesting complexes, where different pathways lead to signals that overlap in 2D spectra.\cite{Turner2012,Engel2007,Panitchayangkoon2010}  
In such instances the ability of femtosecond laser pulses to excite everything within their broad spectral bandwidth can become a limitation rather than a strength. In this work we have utilised a two-colour four-wave mixing experiment to selectively excite the coherence pathways in light-harvesting complexes that have generated such interest. With the enhanced spectral resolution on the excitation pulses it becomes possible to reveal structure within the excited state manifold that cannot be resolved in broadband multidimensional spectroscopy.  Polarization control of these pulses allows further insight into the nature of the coherences that are excited.

We utilise these techniques to explore the coherent dynamics that have been observed in the Phycocyanin-645 (PC645) light-harvesting complex from the Cryptophyte algae, \textit{Chroomonas} CCMP270.  The broadband 2D spectrum from this complex reveals an oscillating cross peak in the region around ($\omega_\tau, \omega_T$)= (519~THz, 499~THz), equivalent to (2.153~eV, 2.071~eV) \cite{Turner2012}.  Oscillations with frequencies 26~THz (108~meV) and 21~THz (87~meV) have been reported and, based on comparisons between rephasing and non-rephasing pulse orderings, attributed to vibrational and possibly electronic coherences, respectively \cite{Turner2012}. 

To specifically excite these coherences the energies of the first two pulses were initially set to be $E_1$=2.171~eV and $E_2=$2.066~eV, which may excite a coherence between states separated by 105$\pm20$~meV.  The third pulse is set to $E_3=2.066$~eV, the same as the second pulse, and interacts with the coherence to give a signal with energy $E_s=-E_1+E_2+E_3$ in the phase matched direction $\mathbf{k_4}=-\mathbf{k_1}+\mathbf{k_2}+\mathbf{k_3} $.  This final interaction can be thought of as a Raman-like process, where the third pulse scatters off the electronic or vibrational coherence.  In this case, however, the final interaction is likely to involve an excited state absorption pathway, based on the large negative peak at this cross-peak location in the broadband 2D spectrum\cite{Turner2011,Turner2012}.  

Regardless of the precise nature of this final interaction, by controlling the delay between the second and third pulses we are able to probe the evolution of the coherence established by the first two pulses. In these experiments the signal energy is given by adding the coherence frequency (negative in this case) to the energy of the final pulse. Conversely, by subtracting the energy of the third pulse from the emission energy, the coherence energy, labelled $\Delta$ can be determined.  

Previous work on PC645 with similar experiments showed extended signal and a decoherence lifetime of 540~fs at 77~K \cite{Richards2012a,Richards2012}. These results also showed oscillations at $\approx$ 22~meV, 40~meV and 74~meV \cite{Richards2012}, providing evidence of a more complex energy structure and strong interactions between electronic and vibrational degrees of freedom, leading to coherent coupling between states not directly excited by the laser pulses.  Conclusive evidence of the nature of the coherence has, however, remained elusive.  In order to gain greater insight into these coherences and beating we consider the results from experiments where the energy of the ${E_{1}}$ pulse was varied between  2.171~eV and 2.217~eV while the energy of the $ E_2$ and $E_3 $ pulses remain constant at 2.066~eV.

\begin{figure}

\begin{subfigure}[ht]{0.45\textwidth}
		\caption{Energy $E_{1}$ = 2.171ev}
		\includegraphics[width=\textwidth]{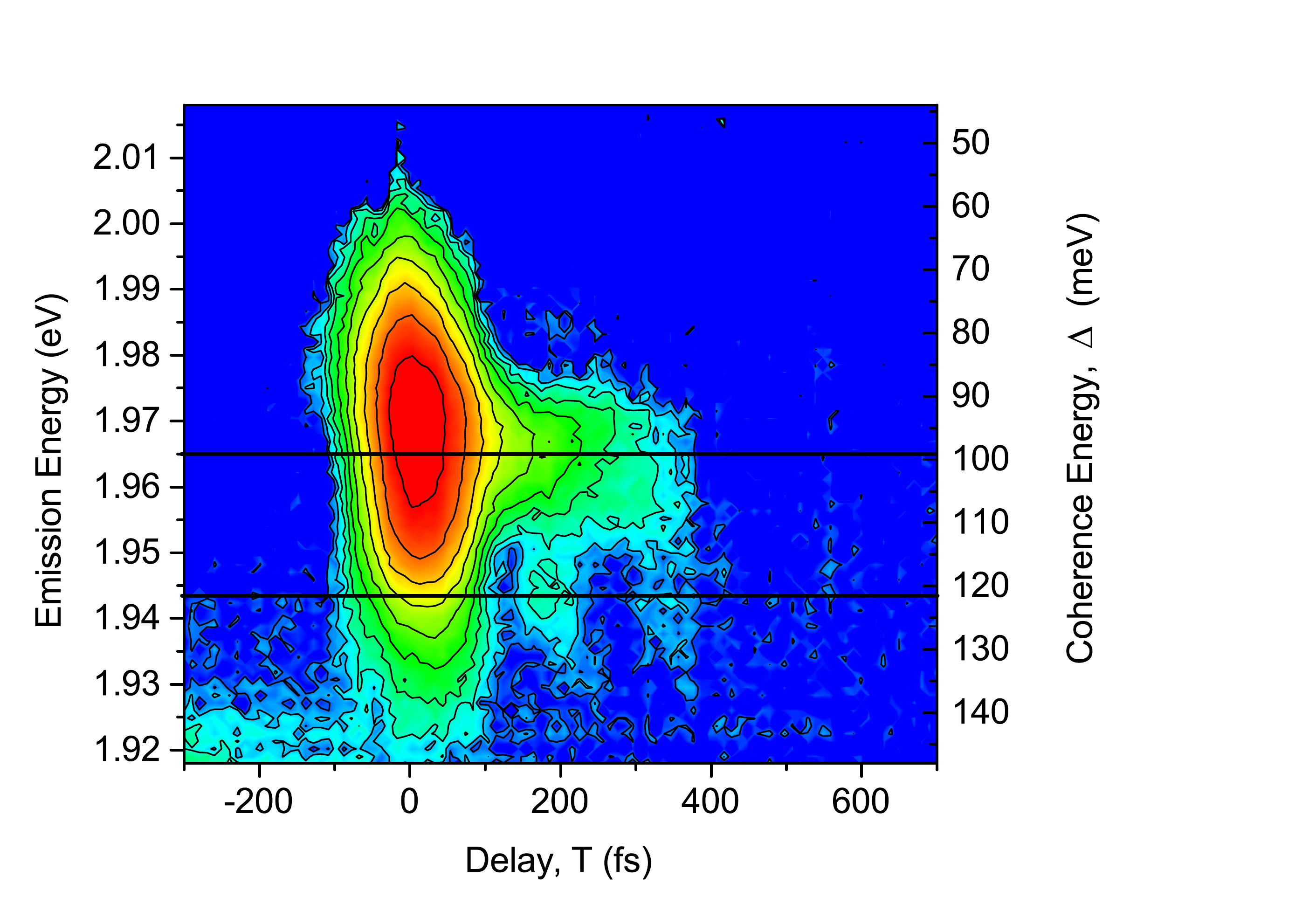}
				\label{w572}
\end{subfigure}
\hfill
\begin{subfigure}[ht]{0.45\textwidth}
		\caption{Energy $E_{1}$ = 2.183ev}
	\includegraphics[width=\textwidth]{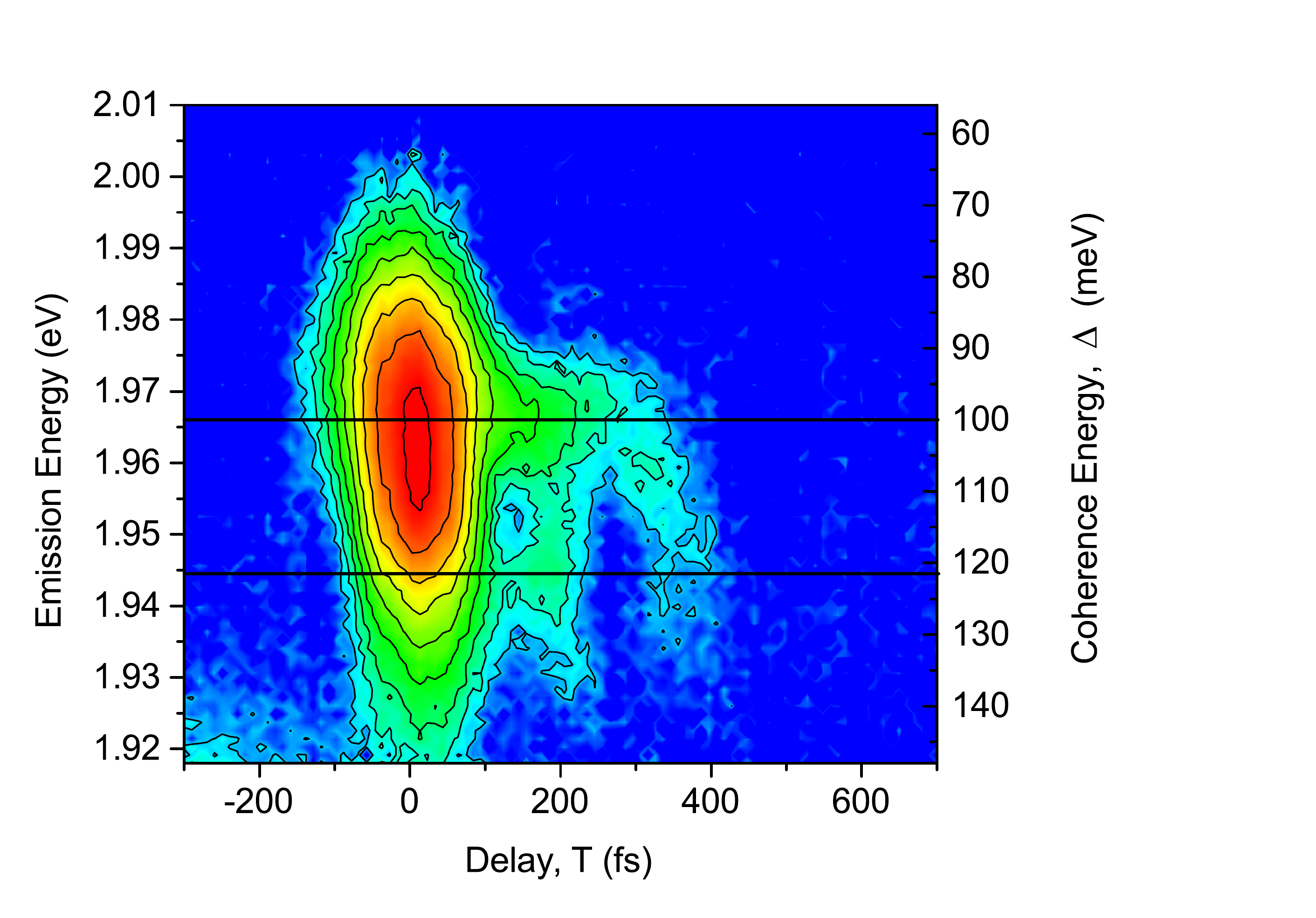}
				\label{w569}
\end{subfigure}
\hfill
\begin{subfigure}[ht]{0.45\textwidth}
		\caption{Energy $E_{1}$ = 2.194ev}
	\includegraphics[width=\textwidth]{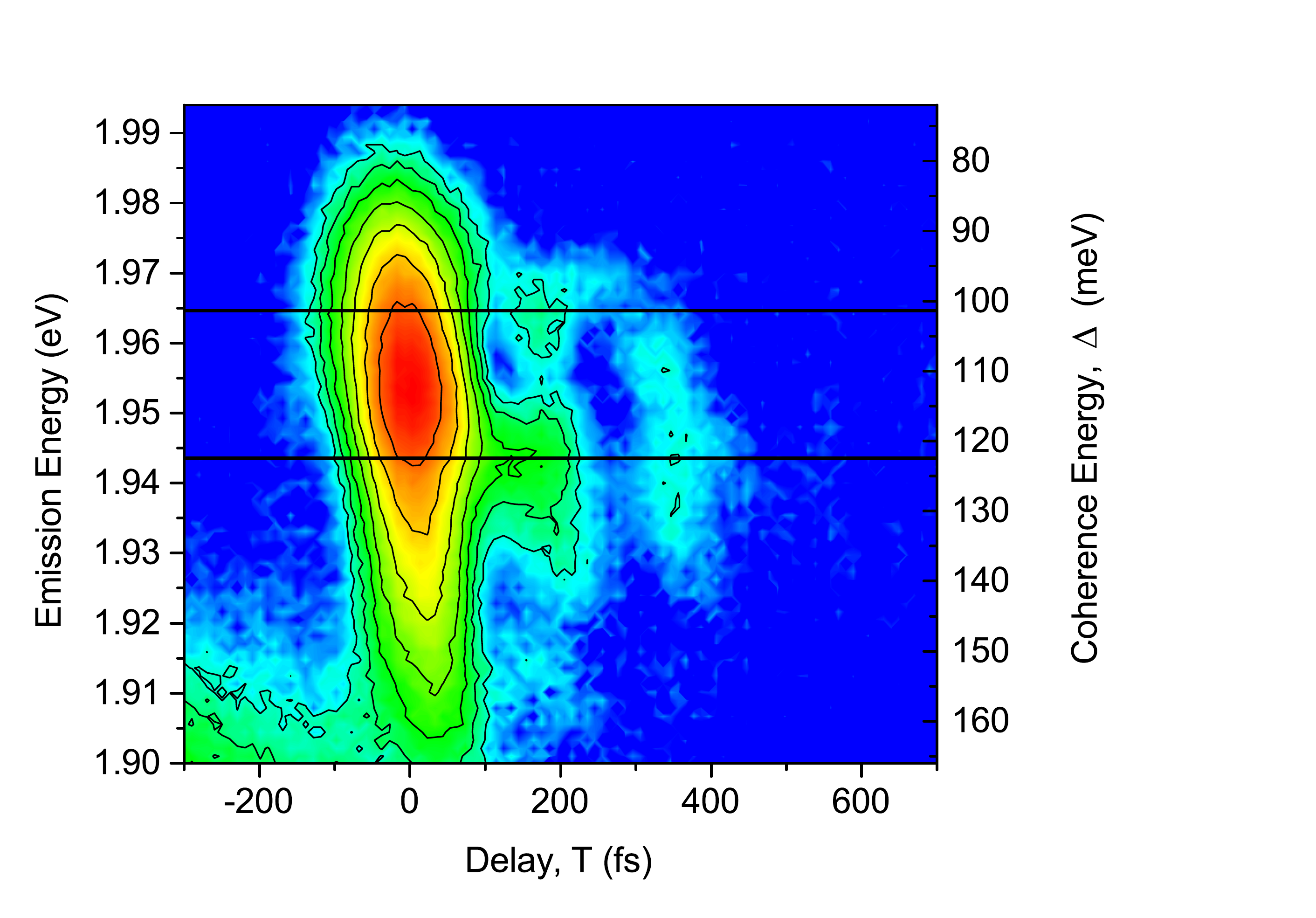}
				\label{w566}
\end{subfigure}
\hfill
\begin{subfigure}[ht]{0.45\textwidth}
		\caption{Energy $E_{1}$ = 2.206eV}
	\includegraphics[width=\textwidth]{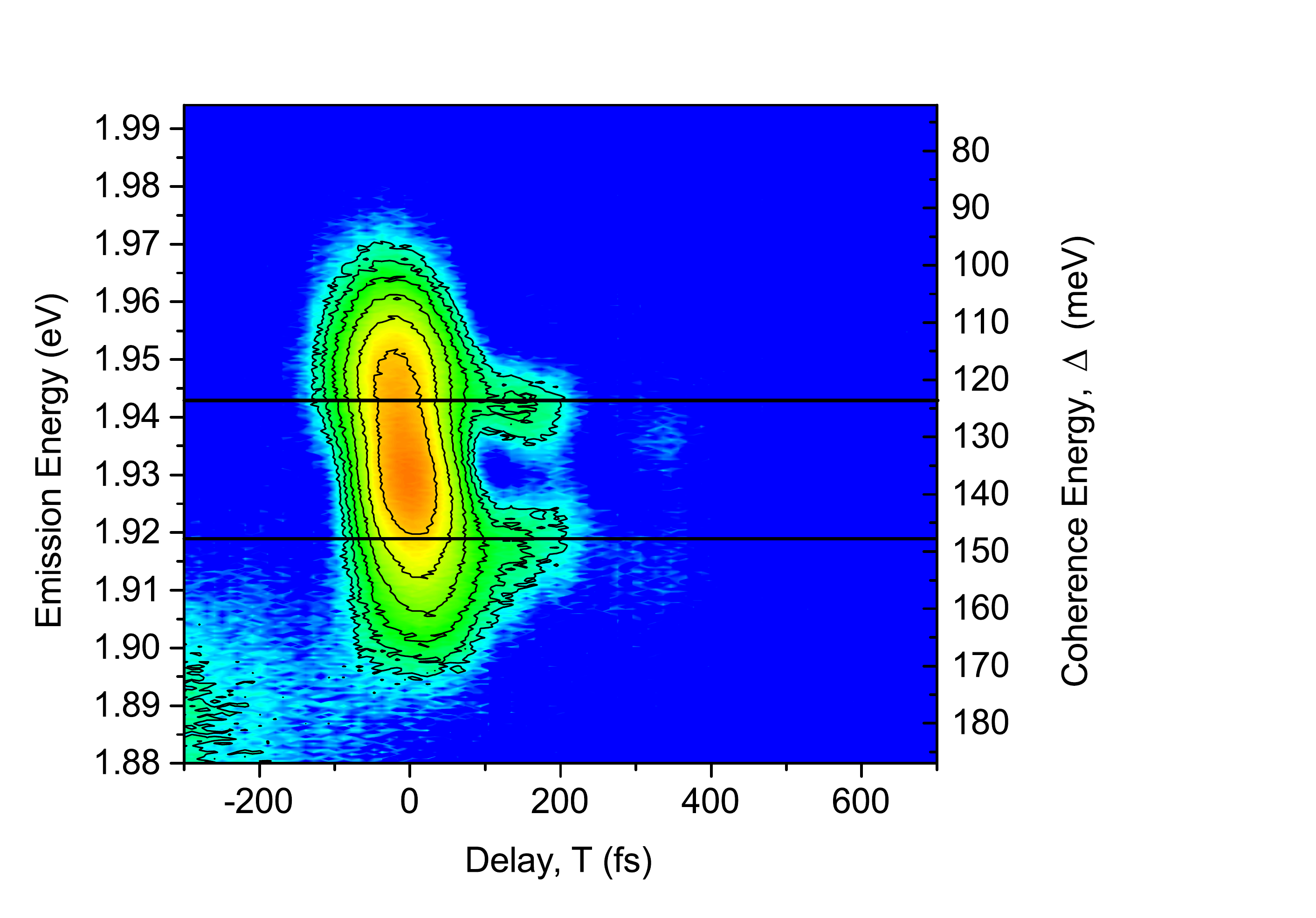}
				\label{w563}
\end{subfigure}
\hfill
\begin{subfigure}[ht]{0.45\textwidth}
		\caption{Energy $E_{1}$ = 2.217eV}
	\includegraphics[width=\textwidth]{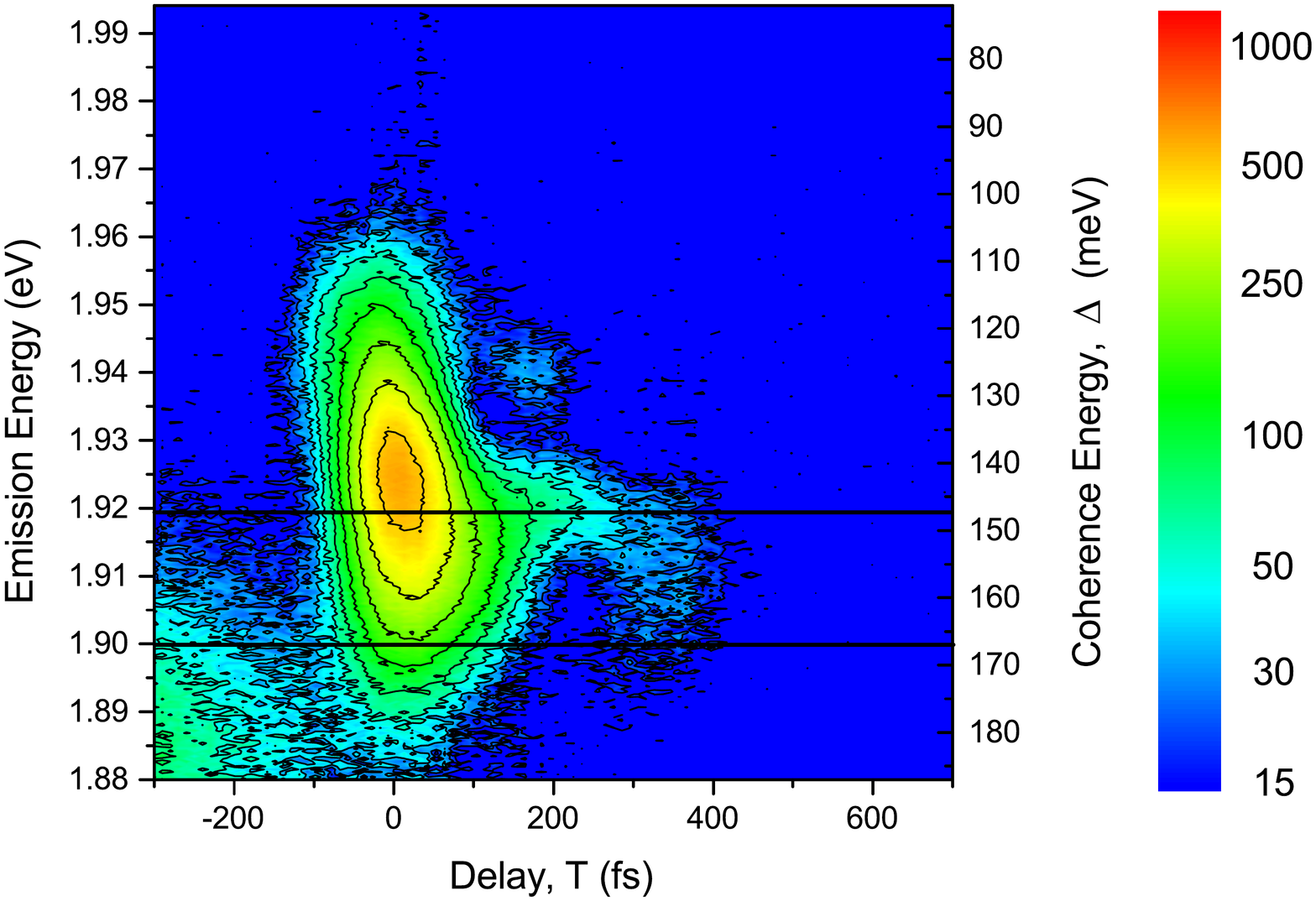}
				\label{w560}
\end{subfigure}
\hfill
\caption{The spectrally resolved intensity as a function of delay, T, for excitation energies $E_{2} = E_{3}=$2.066 eV and $E_{1}$ varied from 2.171~eV to 2.217~eV.  The equivalent coherence energy, $\Delta$ is shown on the right hand axes and the black lines mark the extended coherences.  As $E_{1}$ is increased different coherences are brought into resonance, with a uniform energy separation of ~23meV.}
\label{wldata}

\end{figure}

In Fig. \ref{w572} an extended peak that persists well beyond the pulse overlap region can be seen at an energy of 1.966 $\pm$ 0.002~eV, corresponding to a coherence energy $\Delta =$ 100 $\pm$ 4 meV.  As the energy of the $E_{1}$ pulse is increased this peak remains present in the same position until it is no longer resonant with the energy difference between the first two pulses. An additional peak at 1.943 $\pm$ 0.002 eV emerges  in Fig. \ref{w569} and more clearly in Fig. \ref{w566}, where the $E_{1}$ pulse is set to 2.194~eV.  This peak corresponds to a coherence energy $\Delta =$ 123$\pm$4 meV.  When both peaks are clearly visible, as in Fig. \ref{w569}, oscillations in the amplitude of both peaks can be seen, with a period of 180 $\pm$ 20~fs. This beat period corresponds to an energy difference of $23\pm3$~meV, in close agreement with the energy difference between the two peaks, indicating that the states involved in generating these two coherent superpositions are also coherently coupled.

As $E_{1}$ is increased further, another peak with extended signal at 1.919 $\pm$ 0.002~eV  begins to emerge in Fig. (\ref{w563}) and then possibly a fourth peak at 1.898 $\pm$ 0.002~eV in Fig. \ref{w560}, corresponding to coherence energies of $\Delta =$ 147 $\pm$ 4~meV and $\Delta =168\pm4$~meV, respectively.  These four spectral peaks represent a ladder of states with a uniform energy gap of $23\pm4$~meV.  The beating observed in Fig. \ref{w569} is also present on the other peaks whenever two or more peaks are present, although the amplitude of these oscillations decreases as the coherence energy is increased.  This constant energy spacing between the four peaks and the observation that they are all coherently coupled suggests that we are seeing a ladder of vibrational states for a given vibrational mode.

One further observation in Fig. \ref{wldata} is that each of the signal peaks tend to shift towards greater coherence energy as the delay, T, is increased.  This is seen consistently across many data sets and will be discussed towards the end of this manuscript. 

Our assignment of this vibrational mode at 23$\pm$4~meV is further supported by previous work where an isotropic vibrational mode at this energy was reported in the related C-Phycocyanin(CPC) complex\cite{Womick2012}.  Similar modes have also been found in the traces of transient grating experiments on Allophycocyanin (APC) at 185-216~cm$^{-1}$ (23-26~meV) \cite{Zhang2001} and 205~cm$^{-1}$ (25~meV) \cite{Womick2009}, and again assigned to isotropic vibrational modes. 

\begin{figure}[ht]
          \centering
  		\includegraphics[width=0.8\textwidth]{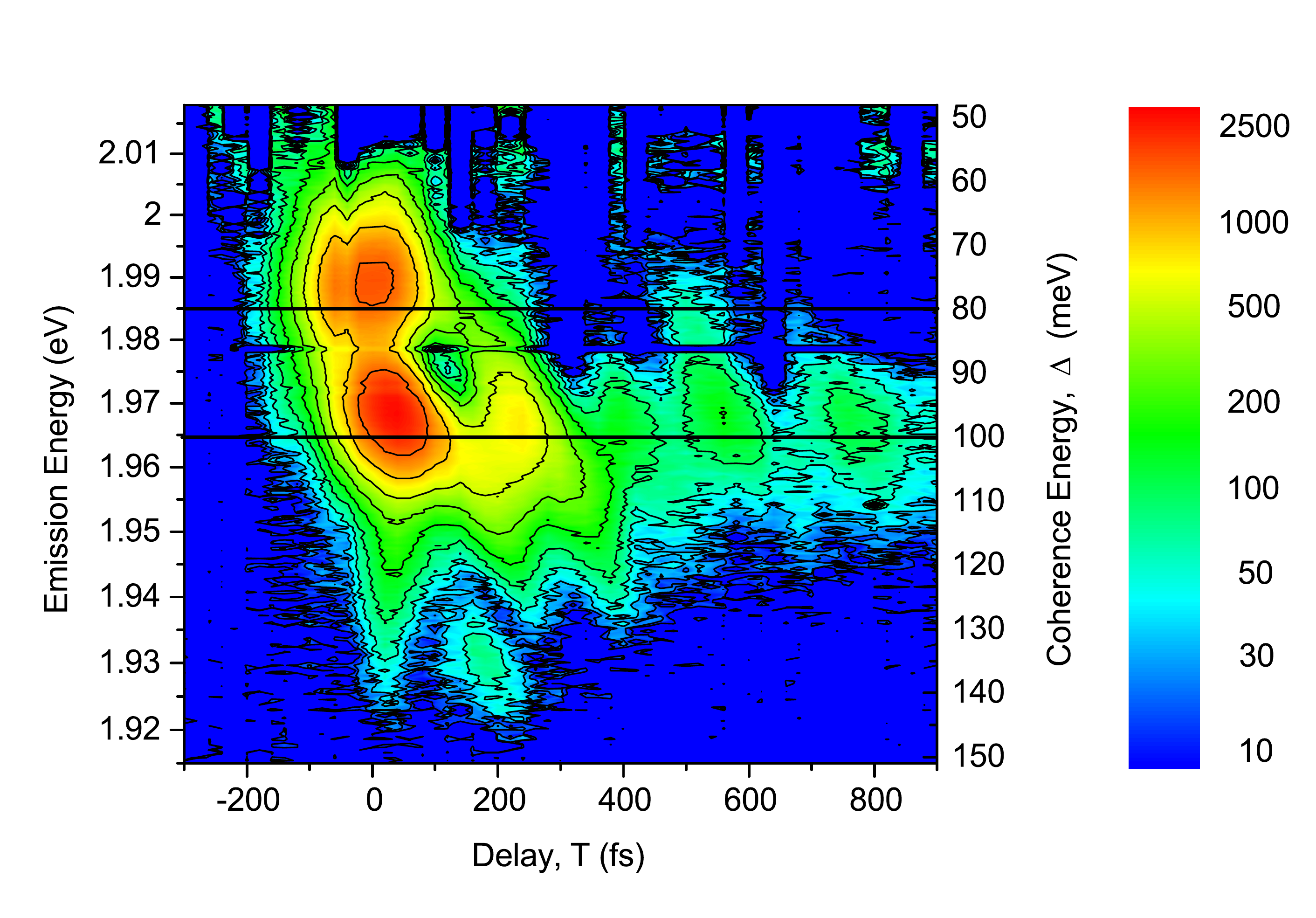}
  		\caption{The measured signal when $E_{1}$ = 2.156~eV and $E_{2} = E_{3}=$2.066~eV. In addition to the long-lived coherence at $\Delta=$100~meV  another, shorter lived,coherence is evident at $\Delta$ = 80 meV.}
  		\label{p576}
  \end{figure}

This vibrational mode and the progression we observe appear to occur on top of a coherent superposition of states that are further apart in energy.  In the results presented thus far the smallest coherence energy for which extended coherences were observed was 100 meV. This underlying coherence could be the fourth level of the 23~meV mode, another vibrational mode or electronic coherence.  To resolve which of these possibilities is most likely we extend this set of experiments by reducing the energy of the $E_{1}$ pulse to 2.156~eV, while keeping $E_2=E_3=$2.066~eV, giving an energy difference between pulses of 90 meV.  In Fig. \ref{p576} the extended coherence at 100 meV remains unchanged and an additional peak at 1.985$\pm$0.003 eV is apparent, corresponding to a coherence energy of 81 $\pm$4 meV. This peak, however decays within 300 fs, compared to $\approx$900 fs for the peak at 100 meV.  This substantially different lifetime suggests that the states involved and possibly the nature of these two coherences are different.  In contrast, the previous experiments increasing the energy of $E_{1}$ greater than 2.171 eV showed the peaks separated by the 23$\pm$3 meV all having roughly the same lifetimes, suggesting the involvement of similar states. 

The much longer coherence time for the 100 meV peak is consistent with vibrational coherence times rather than electronic coherences.  Furthermore, the 100 meV energy difference between the states involved is consistent with previous work that has identified a HOOP (Hydrogen out of plane) wagging mode at 815 cm$^{-1}$ (101~meV) in the related C-phycocyanin\cite{Womick2012} and the work of Turner \textit{et al.}\cite{Turner2012}, who identified a coherence at 26~THz (108~meV) that they attributed to a vibrational coherence based on rephasing and non-rephasing 2D spectra.  

Conversely, the peak at 81 $\pm$ 4 meV is much shorter lived, making it more likely that there is some electronic part to this coherence, and  more closely matches the 21THz (87meV) peak observed in rephasing but not non-rephasing spectra reported by Turner \textit{et al.}\cite{Turner2012}.

\begin{figure}
  \begin{subfigure}[ht]{0.68\textwidth}
  	\centering
  	  	\caption{Polarization Scheme (0,0,0,0)}
  		\includegraphics[width=\textwidth]{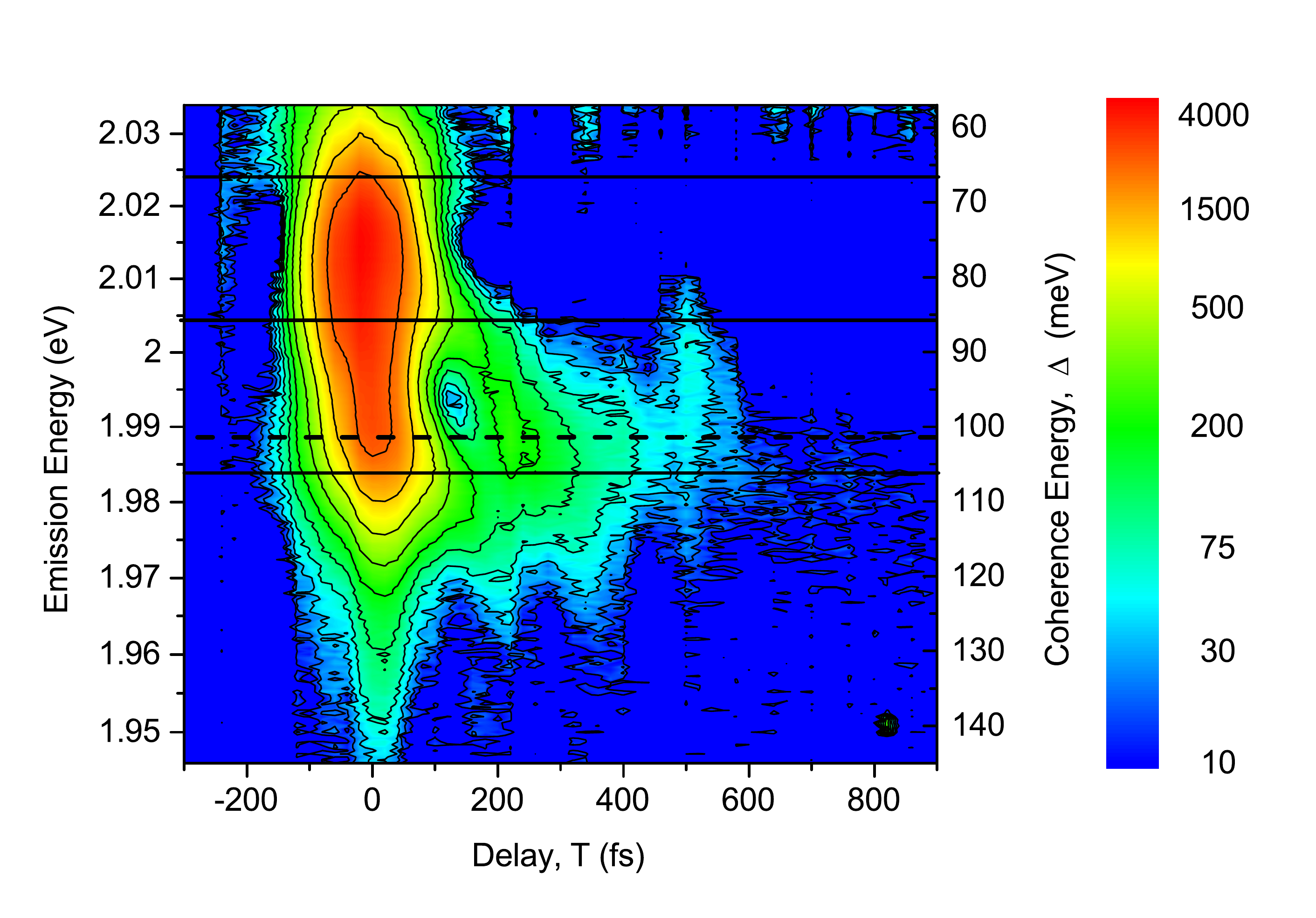}
  		\label{p594}
  \end{subfigure}
  \hfill
 \begin{subfigure}[ht]{0.68\textwidth}
  	\centering
  	\caption{Polarization Scheme  (0,$\pi/2$,$-\pi/4$,$\pi/4$)}
  		\includegraphics[width=\textwidth]{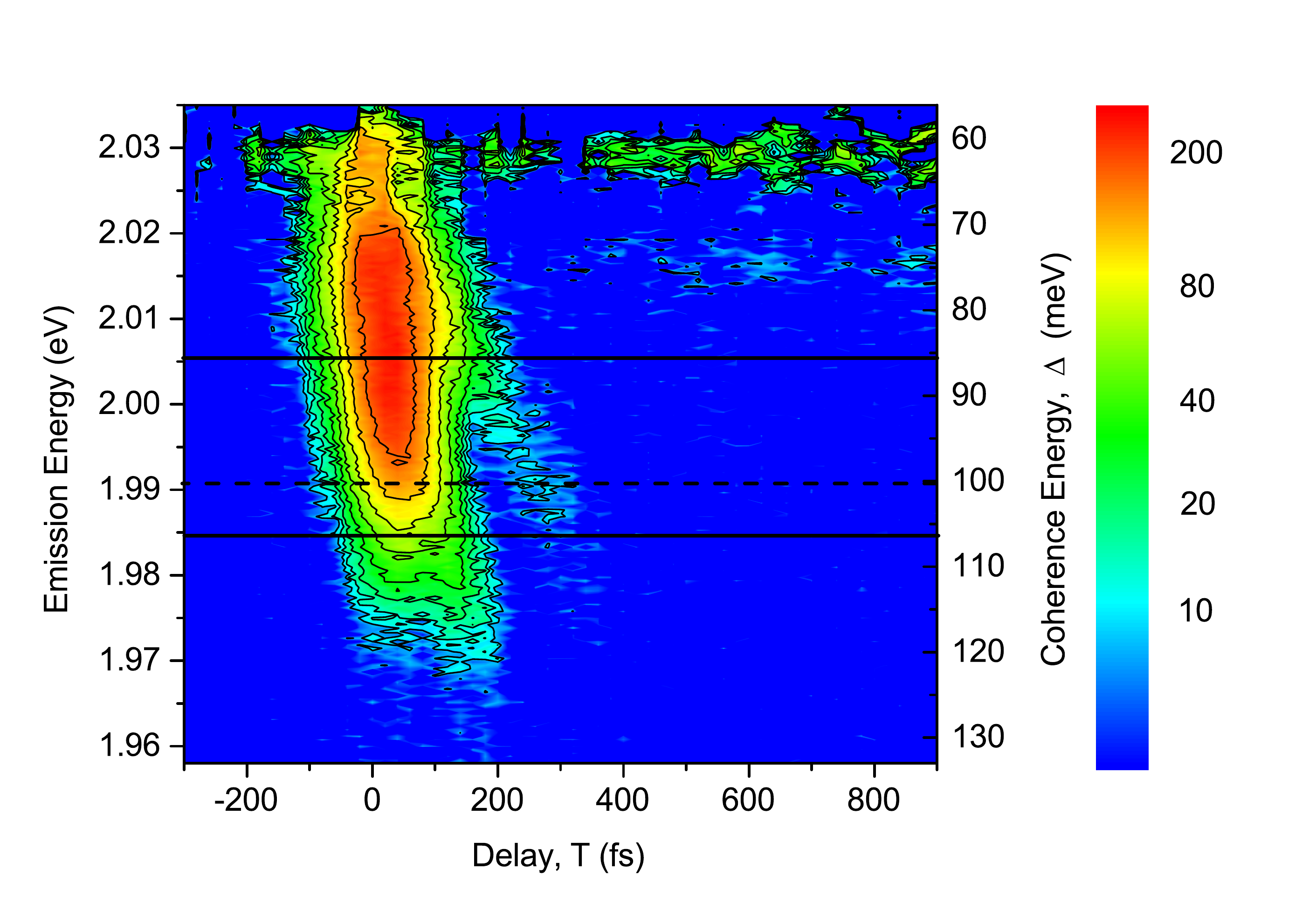}
  		\label{x594}
  \end{subfigure}
  \caption{The measured signals for pulse energies (${E_{1}}$, ${E_{12}}$, ${E_{3}}$) = (2.179 eV, 2.091 eV, 2.091 eV) with (a) all pulses polarized parallel and (b) the cross-polarized scheme described in the text. In (a)  extended signals at $\Delta=$66~meV, 86~meV and 106~meV (black solid lines) are seen in addition to the long-lived signal at $\Delta=100$~meV (dashed line). In (b) the long-lived signal is eliminated but extended coherences at $\approx86$~meV and $\approx106$~meV remain present, indicating coherence between non-parallel transition dipoles. The reduced signal strength in (b) also makes the laser scatter (such as at $\Delta=45$~meV)appear more significant}
  \label{594all}
 \end{figure}
 
It is a feature of electronic states in light harvesting complexes that they are strongly broadened, as can be seen in both linear and 2D spectra previously reported\cite{Turner2012,Collini2010}. We utilise this point to further explore this 81 $\pm$ 4 meV coherence by changing the energies of the excitation pulses to be (${E_{1}}$, ${E_{2}}$, ${E_{3}}$) = (2.179 eV, 2.091 eV, 2.091 eV), which can again lead to the generation of coherent superpositions of states separated by 66-106 meV. Figure \ref{p594} shows that the long-lived signal at a coherence energy of 100 meV is once again present, but in addition, shorter-lived coherences at 66 meV, 86 meV and 106 meV are also observed, again suggesting a different origin.  Further experiments where $E_1$ was varied in this region show the long-lived 100~meV peak to remain unshifted, while the other peaks shift together due to the changing excitation spectra, again indicating that these involve different states to the 100~meV vibrational coherence.

In order to more definitively determine the nature of these states we performed polarization controlled experiments with these pulse energies.  The polarisation of each of the three excitation beams in the experiment can be controlled independently and a polariser in the signal beam path was used to select the signal polarization.  Specifically, we compare the data taken for the case of all beams parallel, with a crossed polarization scheme having polarization angles for the three pulses and signal ($ \theta_1, \theta_2, \theta_3, \theta_4$) = (0,$\pi/2$,$-\pi/4$,$\pi/4$). 

In the crossed polarization scheme used here signal contributions will be eliminated when both the first and second, and the third and fourth light - matter interactions involve transition dipoles that are parallel.\cite{Dreyer2003}  For example, this will be the case when all interactions involve the same two electronic states but different vibrational levels.  In this case, purely vibrational coherences, whether on a ground or excited electronic state, will not give any signal.  Conversely, if the coherences involve different electronic states with transition dipole moments that are not parallel (when considering transitions to the ground state), then we would expect to see some signal.  The strength of the signal will vary depending on the relative angle of the transition dipoles involved.  In the present case, any interactions with electronic states are likely to be with one or both of the excitonic states of the DBV dimer\cite{Mirkovic2007,Collini2010}.  Within the PC645 complex the DBVs are aligned almost parallel, which would give transition dipole moments that are also close to parallel, however, the strong coupling between them leads to the creation of exciton states (symmetric and anti-symmetric combinations of the molecular states) and transition dipoles that are closer to perpendicular than parallel \cite{photosyntheticexcitons}. A coherent superposition of these states, therefore, would be expected to generate significant signal in the cross-polarization scheme described here.  The relative contribution for different pathways and polarization schemes has been calculated as a function of the transition dipole angle and is included in the supporting information.   

The results in Fig. \ref{x594} reveal the absence of the long-lived signal at coherence frequency of 100 meV, but the clear presence of shorter lived signals with coherence frequencies of 86 meV and 106 meV.  The absence of the long-lived coherence at 100~meV confirms that it involves transitions with parallel dipoles and its likely vibrational origin.  The presence of the coherences at 86~meV and 106~meV provides strong evidence that these signals, which persist for >200~fs, arise from coherent superpositions involving different electronic states.  In performing these experiments we have confirmed the polarization angles of each beam to be within $\pm 1\deg$ of the intended angles and the polarization purity to be greater than $ 100:1 $ for each of the pulses.  Based on these numbers and the measured signal in the all parallel polarization case, any signal due to parallel transition dipoles would be expected to be almost an order of magnitude weaker than the signal detected, as shown in the supporting information, confirming that the signal we observe in this case is due to a coherence between states with non-parallel transition dipoles.  A comparison of the measured signal amplitudes in the two cases can give an indication of the angle between the transition dipoles\cite{Ginsberg11}, but separating different signal contributions in the all-parallel case is again a problem and prevents a quantitative comparison.

Based on these observations we have determined an energy level scheme that completely describes all of our observations based on two electronic states and two vibrational modes.  The two electronic states are most likely the two highest energy states of PC645 that have been attributed to the two states of the strongly coupled DBV dimer, labelled DBV+ and DBV-. \cite{Mirkovic2007}

	\begin{figure}[ht]
	        \centering
	        		\includegraphics[width=\textwidth]{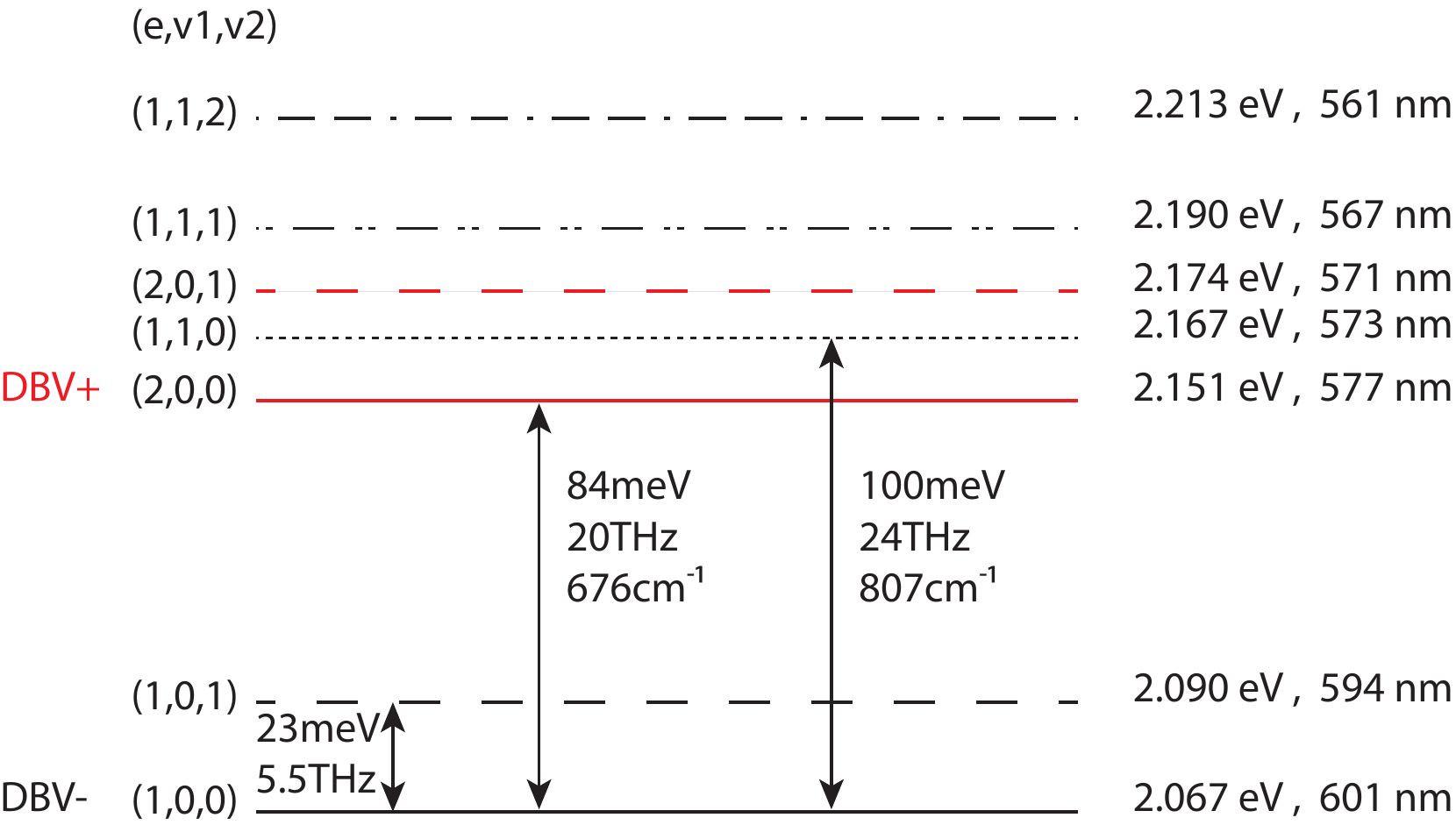}
	\caption{The energy level scheme derived from the experimental results with the two different electronic states labeled black and red, and the different vibrational modes indicated by dotted and dashed lines.}
	\label{Fullenergydiag}
	\end{figure}

From this energy level scheme we can attribute each of the coherent superpositions observed to be either electronic or vibrational in nature.  In all cases, where the coherent superpositions involve the same electronic state but different vibrational levels coherence times >400~fs are observed.  Conversely, where two states involving different electronic states are involved the coherence decays in <300~fs.  These results are not only self-consistent with the data presented here, but also with data obtained with other wavelength combinations and  reported by other groups \cite{Turner2011,Womick2012,Womick2009,Zhang2001,Turner2012} in the same or similar complexes.

While we have represented these states as lines with fixed energy in Fig. \ref{Fullenergydiag} it is important to remember that in reality each state is associated with a complex potential energy surface. Along different points on these surfaces both the absolute energies and the energy differences between states will vary.  Following excitation it is then possible for the excited state wavefunction to move coherently along these potential energy surfaces \cite{Polli}, which may lead to changes in the energy separation between states involved in the coherent superpositions that we measure. Such behaviour would show up in our results as a shift in the coherence energy as the delay, T, is varied. Exactly this type of shift can be seen in much of the data reported here.  Specifically, shifts from low to high coherence energies are observed.  This behaviour is clearest in Fig. \ref{x594} where the elimination of most signal pathways allows the shifting coherence energy to be clearly seen.  The peak that begins at $\Delta=86$~meV appears to shift to $\Delta=100$~meV by 300~fs.  A similar shift of the signal starting at 106~meV can also be discerned.  For the data in Fig. \ref{wldata} a smaller and slower shift in the coherence energy is observed. The difference between the two cases can again be attributed to the different states and hence different potential energy surfaces involved. Where the coherence is between different vibrational levels on the same electronic state the potential energy surfaces will be very similar and any shifts in $\Delta$ will be small and slow to appear.  Conversely, when the two states are different electronic states it is reasonable to expect that the surfaces will differ more substantially and hence $\Delta$ and the energy difference between them will vary more rapidly and by a greater amount. 

Alternative explanations of these apparently shifting signals are possible and to confirm the precise origin further experiments and modelling are required and will be the subject of future work.  We note, however, that this evolution of the coherence frequency would be extremely difficult, if not impossible, to observe in broadband 2D spectroscopy and again highlights the advantages and complementarity  of our two-colour approach.

In summary, we have utilised polarization controlled two-colour experiments to clearly identify coherent superpositions of electronic states in PC645. By combining these observations with experiments where the wavelength of the excitation pulses was varied we have determined a self-consistent energy level-scheme that appears to describe all reported observations of coherences in PC645.  The energy level scheme is based around two electronic states separated by 84$\pm$4~meV, which we attribute to the DBV+ and DBV- states, and two vibrational modes with energy 23$\pm$4~meV and 100$\pm$4~meV. The coherence lifetimes measured are greater than 400~fs when the states involved are different vibrational levels on the same electronic state, and less than 300~fs when two different electronic states are involved. The enhanced spectral resolution that can be achieved with two-colour excitation has made these observations and assignments possible and provided greater clarity regarding coherence among excite states in PC645.  The combination of these techniques with broadband multidimensional spectroscopy provides great potential to further enhance our understanding of coherences, vibrations and the role they play in energy transfer in light harvesting complexes.

\section{Experimental Methods}
PC645 was purified from \textit{Chroomonas sp.}  CCMP270 as described previously\cite{Collini2010}.  The buffer solution with the light-harvesting complexes was diluted 70:30 v/v with glycerol and placed in a quartz cell with pathlength 0.5mm.  The sample was cooled to 77K using an Oxford Instruments cryostat (Optistat).  
A Titanium:sapphire amplifier (Spectra Physics Spitfire) pumped two optical parametric amplifiers (OPAs) that provided the two independently tunable wavelength sources.  The signal emitted in the $k_4=-k_1+k_2+k_3$ direction was spectrally resolved and detected by a 0.27~m spectrometer (SPEX270) and CCD (JY3000).  In each case for the data presented here the delay between the first two pulses was set to 0 fs the delay, $T$, scanned. 

To ensure that each of the excitation pulses were linearly polarized a linear polarizer was included in each beam-path.  To rotate the polarizations a half-wave plate was introduced in two of the beam paths and together with the relevant polarizers rotated to give polarizations at the desired angles.  The purity of the polarization of each beam was checked at the sample position with another linear polariser. The polarization of the signal was determined by placing a linear polarizer at the specified angle directly after the sample. 

\acknowledgement
The authors gratefully acknowledge funding from the Australian Research Council through Discovery Projects.

\begin{suppinfo}
A description of the calculations to determine the orientational factor under different conditions is presented along with plots of this factor for different system properties.  The details of the calculations of uncertainties in the orientational factors and the expected ratio of signals under different polarization conditions are also presented.  Finally, confirmation that the pulses are well compressed and that any extended signal is due to dynamics in the sample is demonstrated in the form of the signal measured in a blank sample with high intensity pulses.
\end{suppinfo}

\bibliography{mybib2}

\end{document}